\documentclass[structabstract]{aa}  
\usepackage{natbib}
\usepackage{graphicx}
\usepackage{txfonts}

\begin{document}
   \title{Sensitivity analyses of  dense cloud chemical models \thanks{The osu.03.2008 network is available in electronic form at the CDS via anonymous ftp to cdsarc.u-strasbg.fr (130.79.128.5) or via http://cdsweb.u-strasbg.fr/cgi-bin/qcat?J/A+A/.}}

   \subtitle{ }

   \author{ V. Wakelam\inst{1,2}, E. Herbst\inst{3}, J. Le Bourlot\inst{4}, F. Hersant\inst{1,2}, F. Selsis\inst{1,2}, S. Guilloteau\inst{1,2}}

   \institute{       Universit\'e de Bordeaux, Observatoire Aquitain des Sciences de l'Univers, BP89 33271 Floirac Cedex, France 
   \and
      CNRS, UMR 5804, Laboratoire d'Astrophysique de Bordeaux, BP89 33271 Floirac Cedex, France 
         \and Departments of Physics, Astronomy, and Chemistry, The Ohio State University, Columbus, OH 43210 USA
         \and Observatoire de Paris, LUTH and Universit\'e Paris-Diderot, Place J. Janssen, 92190 Meudon, France 
              \\
             }

   \date{Received xxx, xxxx; accepted xxx, xxx}

 
  \abstract
   {Because of new telescopes that will dramatically improve our knowledge of the interstellar medium, chemical models will have to be used to simulate the chemistry of many regions with diverse properties.  To make these models more robust, it is important to understand their sensitivity to a variety of parameters.   }
   {In this article, we report a study of the sensitivity of a chemical model of a cold dense core, with homogeneous and time-independent physical conditions, to variations in the following parameters: initial chemical inventory, gas temperature and density, cosmic-ray ionization rate,  chemical reaction rate coefficients, and elemental abundances.}
   {We used a Monte Carlo method to randomly vary individual parameters and groups of parameters within realistic ranges. From the results of the parameter variations, we can quantify the sensitivity of the model to each parameter as a function of time.    Our results can  be used in principle with observations to constrain some parameters for different cold clouds.  We also attempted to use the Monte Carlo approach with all parameters varied collectively.}
   {Within the parameter ranges studied, the most critical parameters turn out to be the reaction rate coefficients at times up to $4 \times 10^{5}$ yr and elemental abundances at later times.  At typical times of best agreement with observation, models are sensitive to both of these parameters.  The models are less sensitive to other parameters such as the gas density and temperature.       }
   {The improvement of  models will require that the uncertainties in rate coefficients of important reactions  be reduced.  As the chemistry becomes better understood and more robust, it should be possible to use model sensitivities concerning other parameters, such as the elemental abundances and the cosmic ray ionization rate, to yield detailed information on cloud properties and history. Nevertheless, at the current stage, we  cannot determine the best values of all the parameters simultaneously based on purely observational constraints.}


     \titlerunning{Sensitivity analyses}
     \authorrunning{Wakelam  et al.}

   \maketitle
%

\section{Introduction}

Molecules are powerful observational tools for the study of the physical conditions and dynamics of star-forming regions. Each step of the formation of a stellar and planetary system is characterized by a chemical composition, which directly reflects the physical conditions of the medium and its evolutionary stage. The far-infrared space telescope Herschel and (sub)-millimeter interferometer ALMA promise to open new windows on the wealth of information found in the interstellar medium (ISM).  With the improvement of observational instrument sensitivity and resolution,  and the opening of new wavelength ranges, more molecules will be detected in the ISM and many molecules will be detected under a wider variety of physical conditions.  The more molecules detected and the more diverse physical conditions  discovered, the more complex chemical models will have to be to reproduce observations.

 The results of chemical models, or simulations,  depend on a number of parameters or groups of parameters often poorly constrained. In this paper, we will be concerned with the sensitivity of simple gas-phase chemical simulations of cold dense interstellar cores to these parameters.  The models are based on the  pseudo-time-dependent approximation, in which the chemistry evolves under fixed and homogeneous density and temperature.  Homogeneous models are also referred to as 0D (zero-dimensional).   Although more physically reasonable models should include the formation of cold cores from more diffuse material as the chemistry progresses, such models are rare given the lack of unanimity of how the collapse proceeds.  

Among the parameters in pseudo-time-dependent/0D models determined only to a limited extent by observations or their comparisons with model results are (1) the gas kinetic temperature and density, (2) the cosmic-ray ionization rate,   (3) the elemental gas-phase abundances, and (4) the initial chemical inventory.    We add to these parameters the rate coefficients (5) of the many chemical reactions despite the fact that they are best determined by laboratory experiments rather than by observational constraints.  
We discuss the parameters in turn.

\begin{enumerate} 

\item The gas kinetic temperature and density are usually determined by the excitation conditions of observed molecular emission lines.  In addition to the uncertainties in the radiative transfer analysis of the lines, inhomogeneities of the gas may exist along the same line of sight.  

 \item There is no direct way to measure the cosmic-ray ionization rate $\zeta$ in dense clouds.  There are two different approaches to constrain $\zeta$.  The first one, usually used, is to determine it by comparison of modeled and  observed abundances for specific ionic species or their ratios.  Currently, there appears to be a dichotomy between diffuse clouds, in which the high abundance of H$_{3}^{+}$ leads to a large value of $\zeta$, and dense clouds, in which the ionization conditions often lead to a value of $\zeta$ 1-2 orders of magnitude lower  \citep{1982ApJ...255..160W,1998ApJ...499..234C,2003Natur.422..500M, 2004A&A...417..993L,2009ApJ...694..257I}.  There is some indication that the difference is simply due to the inability of low energy cosmic rays to penetrate into the center of dense clouds \citep{2009A&A...501..619P}.  A second approach is to use the value measured in the solar system \citep{1968ApJ...152..971S,1998ApJ...506..329W} and consider that $\zeta$ is constant in the Galaxy. However, in the solar system, the low energy end of the cosmic ray spectrum cannot be determined directly because of the solar wind.  

\item The computed abundances of atomic and molecular species depend on the choice of elemental abundances in the gas phase. Elements are measured in absorption in the diffuse medium \citep{2004oee..symp..336J}. In order to reproduce the observed gas-phase composition of dense clouds, it is often assumed that heavy atoms, including S, Si, Fe, Na and Mg, are depleted further from the gas between the diffuse (or translucent) and dense phases of a cloud life. Specifically,  the initial abundances of these elements in pseudo-time-dependent models are taken to be 2 to 3 orders of magnitude lower than in diffuse clouds but the efficiency of the depletions onto the grains is quite uncertain \citep{1982ApJS...48..321G}.  Moreover, the depletions of the elements may depend on the depth into the molecular cloud.

\item Since the $t=0$ time of pseudo-time-dependent models is artificial, the initial chemical inventory chosen can be artificial as well.  Chemical models usually start with all the elements in the atomic form, except for H$_{2}$, assuming that the density of the gas had suddenly increased from a diffuse to a dense medium.   Shock models of the conversion of diffuse to dense gas show, on the other hand, that the process is slow and that a significant amount of CO may be synthesized before a sizable dense cloud core is produced \citep{2004ApJ...612..921B}.   Preliminary results show that the formation of complex molecules is far slower if the starting form of carbon is CO (Hassel, Herbst, \& Bergin, in preparation). Thus, the initial chemical abundances and the cloud age, as determined by comparison of observational and model abundances, are correlated.	

\item Finally, the chemistry in the dense ISM involves thousands of gas-phase reactions, especially when one is dealing with complex molecules. Only a small fraction of these reactions have been studied at the low temperatures  present in cold dense cores.  Thus, even for systems studied in the laboratory at higher temperatures, there can be a difficulty in extrapolating results to significantly lower temperatures.   The most important reactions at assorted cloud ages have been discussed in previous papers with our sensitivity approach and those of others \citep{2004AstL...30..566V,2005A&A...444..883W,2006A&A...451..551W,2008ApJ...672..629V}.

\end{enumerate}

 In order to quantify the reliability of the chemical models and to determine how best to improve them, we need to know how sensitive they are to these parameters so that we can reduce the uncertainties in the more important ones.  For example, if it is found that model results are more sensitive to current uncertainties in chemical rates than to elemental abundances, then future measurements of such rates should rank at high priority.  If the opposite is true, priority might be given to developing better dynamic models of clouds to follow the depletion of elements from the gas more appropriately.   Moreover, if the models are compared with well-known cold clouds such as TMC-1 and L134N, the observations can be used to further constrain some of the model parameters within their computed uncertainties.  In this article, we present sensitivity analyses of gas-phase cold dense cloud chemistry to all of the parameters discussed above.   Details on the model are given in Section 2, while the sensitivity method and the ranges of values over which the parameters are varied are described in Section 3. The results of these analyses are discussed in Section 4 while the constraints that observations of specific cold cores can yield concerning the model parameters are presented in Section 5.  Finally, Section 6 contains a discussion.  In our sensitivity calculations, one can proceed via the independent variation of parameters or groups of parameters, or one can attempt to vary all parameters collectively.  Although the latter approach is attempted, as discussed below, most of our useful results stem from the former treatment.

\section{Chemical model}

We used a new version of the Nahoon gas-phase chemical code \citep{2005A&A...444..883W}, which is significantly faster than the previous version.  The code is written in Fortran 90 and the differential equation solver is DLSODES from the ODEPACK package (http://www.netlib.org/odepack/opkd-sum). Although the code is written for 1D models (models that can be heterogeneous in one dimension), we used it in 0D with gas temperature and total hydrogen density set at the standard values for cold dense cores of $T=$10~K and $n_{\rm H}$= $2\times 10^4$~cm$^{-3}$. The visual extinction is set to 10 so that photo-chemistry with external UV photons is of little importance. The cosmic-ray ionization rate $\zeta_{\rm H}$ is $1.3\times 10^{-17}$~s$^{-1}$.
We used the osu.03.2008 chemical network (http://www.physics.ohio-state.edu/$\sim$eric/research.html), which contains 13 elements, 455 species, and 4508 gas-phase reactions. This particular version of the osu network does not contain molecular anions, nor any dynamic depletion of gas-phase species.  No reactions on grains are considered except for the production of H$_{2}$ from two H atoms, a process that is treated in a pseudo-gas-phase manner by relating the density of grains to the overall gas density. The osu.03.2008 chemical network used in this work is available at the CDS with the following format. The first seven columns contain the reactants (three columns) and the products (four columns) of the reactions. The next three columns ($\alpha$, $\beta$ and $\gamma$) give the parameters to compute the rate coefficients. Column 11 indicates the type of reaction. Column 12 is the number of the reaction within a given type of reaction whereas column 13 is the general number of the reaction in the network. Finally, the last column gives the uncertainty factor for the rate coefficient. More details about the format can be found at http://www.physics.ohio-state.edu/$\sim$eric. Species are initially all in the atomic form except for H, which is 100\% in H$_2$. Table~\ref{stand_mod} summarizes our standard model.  

Listed in Table~\ref{elem_ab}, the base elemental abundances used are oxygen-rich,  and represent a low-sulfur version of the set labeled EA3 by \citet{2008ApJ...680..371W}. In this set of abundances, the amount of helium  is based on the observed value in the Orion Nebula \citep{1992ApJ...389..305O}. The oxygen, nitrogen and carbon abundances seem to be rather constant in diffuse clouds and we have taken the values observed in $\zeta$ Oph \citep{1993ApJ...402L..17C,1998ApJ...493..222M}. Since we do not include gas depletion in Nahoon, we then removed 2.4\% of C and 13\% of O to account for the fraction depleted on grain mantles in the form of CO and H$_2$O, as suggested by \citet{1995A&A...296..779S}. For Na$^+$, Cl$^+$ and P$^+$, we use the observed values in $\zeta$ Oph \citep{Savage1996}. In dense clouds, iron, silicon and magnesium show additional depletion compared to the diffuse clouds, and we adopted the depleted values recommended by \citet{2003MNRAS.343..390F}.  Sulphur is the most problematic element \citep[see for instance][]{1999MNRAS.306..691R}. The low value, taken from \citet{1982ApJS...48..321G}, is the standard ``low-metal''  abundance.  In summary, our low-sulfur EA3 abundances  contain  probable elemental abundances based on observational constraints. One of the consequences is that we do not produce large abundances of complex molecules since the C/O ratio is significantly less than unity.   These base abundances will be varied to determine the sensitivity of the model to elemental abundances. 

\begin{table}
\caption{Standard dense cloud model.}
\begin{center}
\begin{tabular}{l|c}
\hline
\hline
Parameter & Value \\
\hline
$T$ & 10~K \\
$n_{\rm H}$ & $2\times 10^4$~cm$^{-3}$\\
$A_{V}$ & 10 \\
$\zeta_{\rm H}$ & $1.3\times 10^{-17}$~s$^{-1}$\\
Initial abundances & atomic except for 100\% H$_{2}$ \\
Rate coefficients & osu.03.2008\\
\hline
\end{tabular}
\end{center}
\label{stand_mod}
\end{table}

\begin{table}
\caption{Elemental abundances with respect to total hydrogen nuclei.}
\begin{center}
\begin{tabular}{l|c}
\hline
\hline
Species & Abundance \\
\hline
He & 0.09  \\
N & $7.6\times 10^{-5}$ \\
O & $2.56\times 10^{-4}$ \\
F & $6.68\times 10^{-9}$ \\
C$^+$ & $1.2\times 10^{-4}$\\
S$^+$ & $8.0\times 10^{-8}$  \\
Si$^+$ & $0.0$ \\
Fe$^+$ & $1.5\times 10^{-8}$  \\
Na$^+$ & $2.0\times 10^{-7}$  \\
Mg$^+$ & $0.0$  \\ 
Cl$^+$ & $1.8\times 10^{-8}$ \\
P$^+$ & $1.17\times 10^{-7}$ \\
\hline
\end{tabular}
\end{center}
\label{elem_ab}
\end{table}

\section{Sensitivity analyses}

\subsection{Variations for most parameters}\label{var_par}

\begin{table}
\caption{Variational ranges of the parameters.}
\begin{center}
\begin{tabular}{lll}
\hline
\hline
Parameter & Range & N runs \\
\hline
Reaction rate coefficients & Uncertainty Factor & 2500 \\
Temperature$^*$ & 5-15 K & 2500 \\
H density$^*$ & $(1-3) \times 10^{4}$ cm$^{-3}$  & ---  \\
Elemental abundances &  $\pm$ 50\% & 2000 \\
Cosmic-ray ionization rate ($\zeta$) & $(0.5 -5.0) \times 10^{-17}$ s$^{-1}$  & 1000 \\
Initial Concentrations & see text & 3500 \\
\hline
\end{tabular}
\end{center}
\label{param}
$^*$ Varied together
\end{table}%

The major goal of this analysis is to quantify the sensitivity of chemical models to variations of the parameters within realistic ranges of values. The method consists in generating new sets of parameters ($P_{\rm n}$) using random functions and computing the corresponding chemical fractional abundances vs time ($X_{\rm n}(t)$). Using such an approach, we studied the sensitivities to the reaction rate coefficients, the gas temperature and density, the elemental abundances, and the cosmic-ray ionization rate. The variational range for each parameter is given in Table~\ref{param}.  For the reaction rate coefficients, we have assumed an amplitude variation equal to their uncertainties, as originally included in the UMIST database (see http://www.udfa.net/) and subsequently in the osu database, with a log-normal distribution \citep[see][]{2005A&A...444..883W,2006A&A...451..551W}. We define $F_{\rm i}$,  the uncertainty factor of the rate coefficient, at a 1$\sigma$ level of confidence. As a consequence, there is a   68.3\% chance that the rate coefficient lies between $k_{\rm i}/F_{\rm i}$ and $k_{\rm i} \times F_{\rm i}$, where $k_{\rm i}$ is the listed rate coefficient for reaction $i$.  
For the other parameters, except for the initial chemical inventory (see below), flat distributions (no preferred values) were assumed with ranges determined by reasonable uncertainties, which are mainly $\pm$50\% from the center or peak value. The gas temperature  then lies between 5 and 15~K, the total hydrogen density between $1 \times 10^4$ and $3\times 10^4$~cm$^{-3}$,  and the elemental abundances between $\pm$50\% of their standard values.  The cosmic ray ionization rate, which is clearly uncertain to more than a factor of two, is varied between $5\times 10^{-18}$ and $5\times 10^{-17}$~s$^{-1}$.

The variations of the elemental abundances are sufficient to probe perturbations around the chosen values, including C/O ratios greater than unity,  but not large enough to include the total diversity used by previous modelers.   All individual parameters (e.g., the cosmic ray ionization rate) or groups of parameters (e.g., reaction rate coefficients, elemental abundances)  were varied one at a time except for the temperature and the density, which were varied together although they are not correlated.   In total, we ran 8 000 different models of the dense cloud in varying these parameters. We used a Monte Carlo method to sample parameters randomly; this is most efficient for groups of parameters such as the reaction rate coefficients. For individual parameters such as the cosmic-ray ionization rate, a non-random sampling method could have been used to span the range of values. 

To quantify the dispersion of the computed abundance profile of  a specific species  at a given time when one parameter or a number of different parameters is varied, we calculate the  standard deviation of the profile in logarithmic units using the formula
\begin{equation}
\sigma(t)=\sqrt{\frac{1}{N} \sum_{n=1}^{N}(log(X_n(t)) - \overline{log(X(t))})^2}
\end{equation}
where the sum is over the number of computed abundances $N$, $X_n(t)$ is the abundance of the specific species for run $n$ at  time $t$ and $\overline{log(X(t))}$ is the mean of the logarithm of the abundance at  time $t$. We have chosen to compute $\sigma$ in logarithmic units in order to deal with numbers over many orders of magnitude.  In addition, the uncertainties of most rate coefficients are given as factors rather than additive terms, making logarithmic units more natural.  
 If the abundance profile at  time $t$ follows a log-normal distribution (Gaussian in logarithmic units) with $\sigma(t)=1$, 68.3\% of the abundances lie within a factor of 10 from the mean abundance. In the rest of the paper, we will use this definition of the standard deviation. 

\subsection{Determination of the initial chemical inventory}

\begin{figure}
\includegraphics[width=1\linewidth]{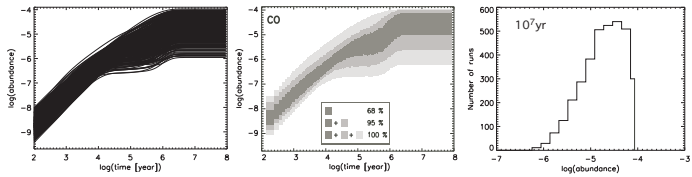}
\caption{CO abundance computed in a diffuse/translucent cloud. Left panel: plot of 10\% of the runs. Middle panel: the gray levels represents the percentage of runs. Any level of gray: at least one run. Two darkest gray levels: 95\% of all runs at one given time (2 sigma if the distribution is log-normal). Darkest gray: 68\% of all runs  (1 sigma if the distribution is log-normal). Right panel: distribution of the fractional abundance at $10^7$~yr, where the median value is approximately $2\times 10^{-5}$. \label{allrun_CO}}
\end{figure}

The initial atomic and molecular abundances for the cold core are treated uniquely. To study the sensitivity to the initial concentrations, we need to have realistic distributions for dense cloud models. Since dense molecular clouds are presumably formed from diffuse or translucent gas, we first compute the steady-state chemical composition of such a gas with a temperature of 50~K, a total hydrogen density of $2\times 10^3$~cm$^{-3}$ and a visual extinction of unity. The species start in their atomic form prior to the diffuse/translucent cloud stage, except for hydrogen, which is assumed to be divided equally among atomic and molecular forms.   The fixed elemental abundances used for the diffuse/translucent stage are the ones listed in Table~\ref{elem_ab}, although a more realistic set would have less depletion.  In order to get realistic distributions, we randomly varied the reaction rate coefficients within their uncertainty ranges using an additional 3500 runs.  The abundances and distributions we obtained at steady-state were then used as the initial chemical inventory for the dense cloud.  As an example, the CO abundance distribution in the diffuse/translucent cloud conditions is shown in Fig.~\ref{allrun_CO} (left and middle panels) as a function of time. The distribution of CO at steady-state ($10^7$~yr) used as the initial distribution  for the dense cloud is shown on the right panel and can be fitted by a log-normal distribution.  It peaks at a fractional abundance  of $\approx 2\times 10^{-5}$ with a standard deviation of a factor of 6.3. For comparison, the C$^{+}$ and C distributions have their steady-state peaks at $\approx 6\times 10^{-5}$ and $\approx 10^{-5}$ respectively, so that the gas is still mainly atomic except for H$_{2}$, which takes up most of the elemental hydrogen; the atomic H distribution peaks at $\approx 3\times 10^{-3}$. These results are somewhat different from our typical assumption that carbon and oxygen are purely atomic and hydrogen purely molecular at the initial stage of a dense cloud. Our distributions span two extreme scenarios: one in which most of the carbon is still in the atomic form (C$^+$ abundance $\approx 1 \times 10^{-4}$) and one in which CO is the dominant form of carbon (CO abundance  $\approx 1 \times 10^{-4}$). These extreme values can arise depending upon parameters in the shock models for dense cloud formation  of  \citet{2004ApJ...612..921B}.


\subsection{Variation of all parameters}
\label{variation}
 
\begin{figure}
\begin{center}
\includegraphics[width=1\linewidth]{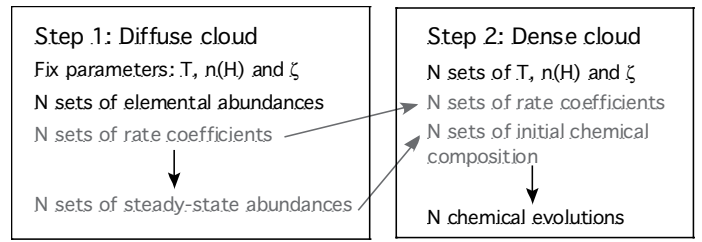}
\caption{Steps used to vary all the parameters of the model collectively. \label{steps}}
\end{center}
\end{figure} 

In the sections presented above, we have emphasized how we vary the parameters or groups of parameters one after the other to study their effect independently.  The alternative approach is to develop a model in which all parameters are varied collectively. To achieve this goal, we have once again used a  two-step procedure involving the diffuse and dense cloud stages, discussed below.   The procedure is also summarized in Fig.~\ref{steps} for better understanding. 

For the first step, we once again compute the steady-state chemical composition of a diffuse cloud with a fixed temperature of 50~K, a density of $2\times 10^3$~cm$^{-3}$, visual extinction of unity,  and cosmic-ray ionization rate of $1.3\times 10^{-17}$~s$^{-1}$ (as described in the previous subsection). But now we vary at the same time the elemental abundances and the rate coefficients within the ranges given in section~\ref{var_par}, using the Monte-Carlo approach with a different distribution of random numbers for both sets of parameters.  We obtain $N=3500$ different steady-state chemical compositions as a consequence of varying the sets of elemental abundances and rate coefficients simultaneously.  In the second step, we use this distribution of 3500 chemical compositions as the initial inventory and compute the chemical evolution of a dense cloud using  the previous 3500 sets of rate coefficients.  In addition, we vary simultaneously the temperature, the density and the cosmic-ray ionization rate using different random number distributions for each parameter.  We then obtain 3500 different chemical evolutionary paths, each of which corresponds to to a different temperature, density,  cosmic-ray ionization rate, set of rate coefficients, set of elemental abundances and initial chemical composition.

\section{Results}

In subsections 4.1-4.4 below, we discuss the sensitivities of the abundances to individual variations of parameters or groups of parameters with the others set equal to their standard values.  Afterwards we compare these sensitivities and also consider the sensitivities to all parameters varied collectively.

\subsection{Sensitivity to the initial concentrations and their distributions}
\label{sens_ic}

\begin{figure}
\begin{center}
\includegraphics[width=1\linewidth]{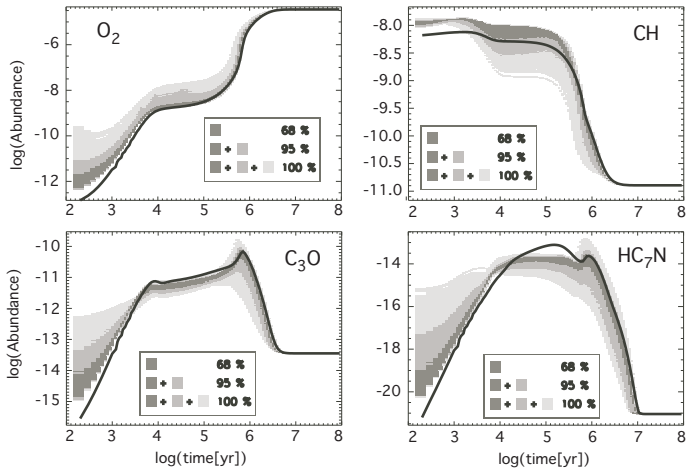}
\caption{O$_2$, CH, C$_3$O and HC$_7$N abundances as a function of time in the dense cloud stage, starting with a distribution of atomic and molecular abundances discussed in the text. 
The black lines represent abundances obtained if we use our standard atomic initial concentrations except for H$_{2}$.  \label{CI_sens}}
\end{center}
\end{figure}

\begin{figure}
\begin{center}
\includegraphics[width=0.47\linewidth]{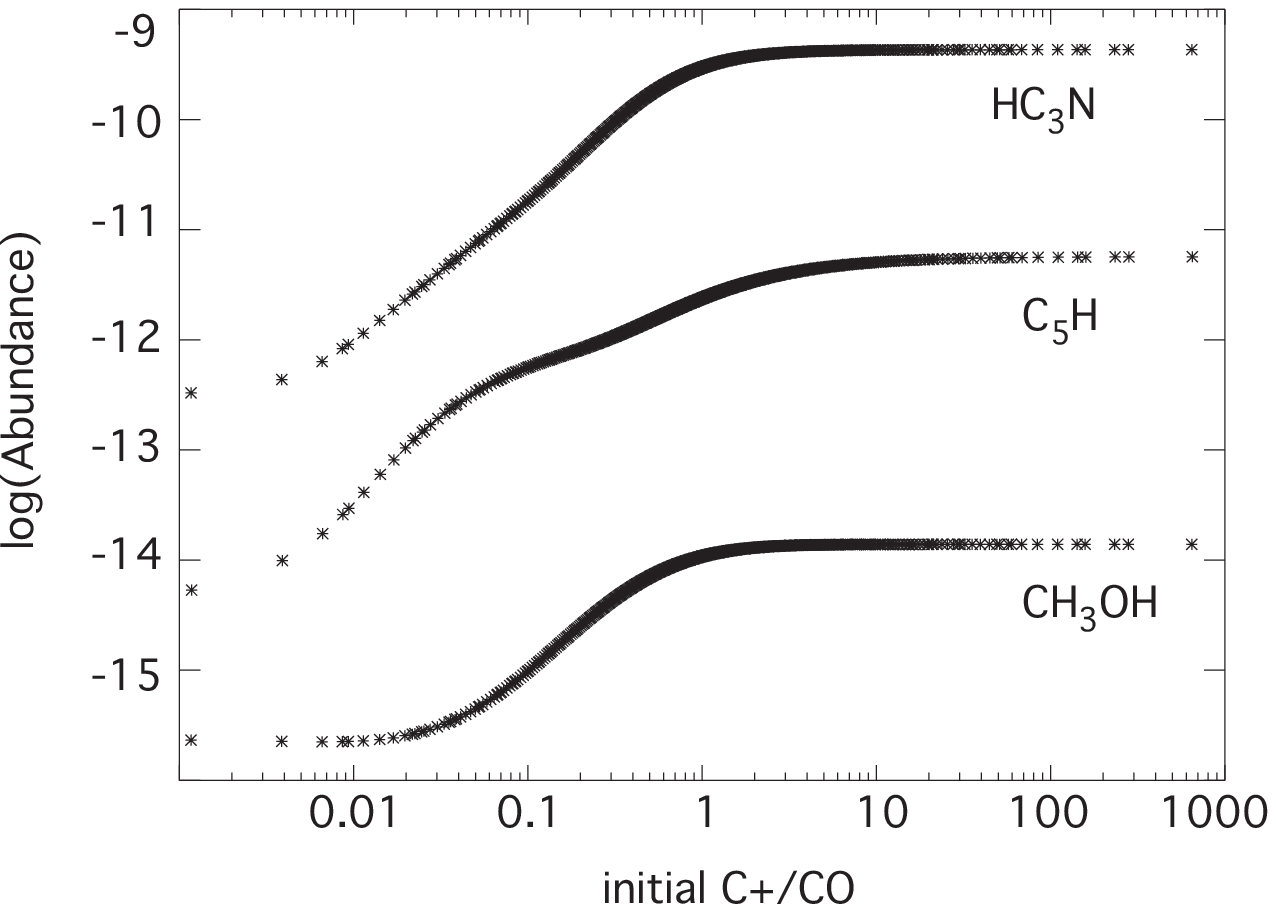}
\includegraphics[width=0.45\linewidth]{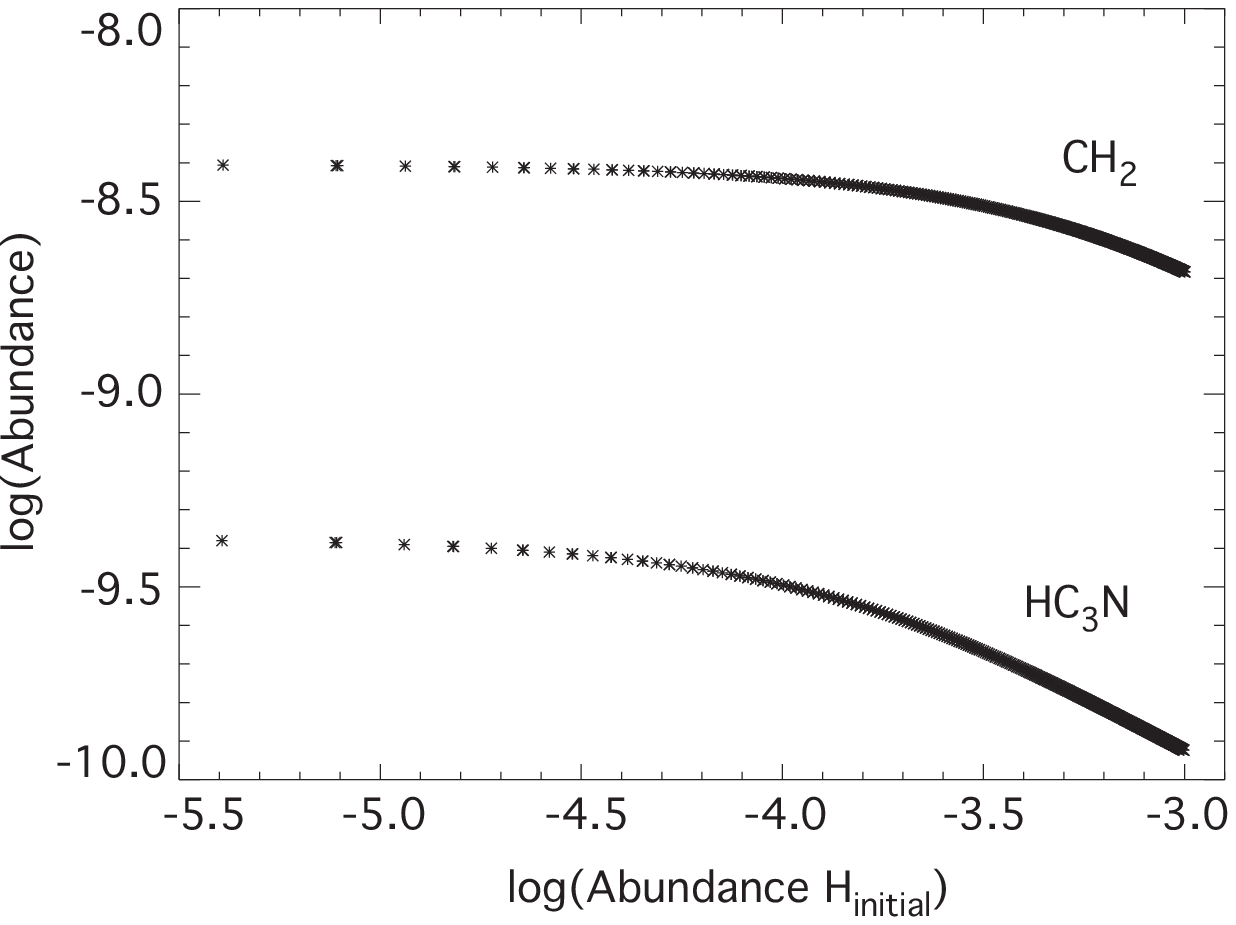}
\caption{Left panel: Calculated HC$_3$N, C$_5$H and CH$_3$OH abundances as a function of the initial concentration ratio  C$^+$/CO of the dense cloud stage. Right panel: Calculated CH$_2$ and HC$_3$N abundances as a function of initial atomic hydrogen fractional abundance. Results are shown for $10^5$~yr. \label{corr_init}}
\end{center}
\end{figure}

The sensitivities of selected dense cloud abundances to changes in initial chemical concentrations  are shown in Fig.~\ref{CI_sens} as functions of time.  The initial distributions of the molecular concentrations introduce spreads in the time-dependent solutions that last until the system is at steady state, where abundances no longer depend on the initial concentrations because we do not have any bistability for the parameters used. Although some distributions, such as that for atomic H, show standard deviations ($\sigma=0.23$, see section~\ref{var_par}) as large as a factor of $\approx 2$ for times in the dense cloud stage up to 10$^{5}$ yr, most distributions are more sharply peaked, as can be seen in Fig.~\ref{CI_sens}.   For the species shown in this figure, the actual logarithmic standard deviations $\sigma$ for O$_2$, CH, C$_3$O and HC$_7$N are 0.14, 0.12, 0.07 and 0.17 respectively at $10^5$~yr. For comparison, we have started the dense cloud stage with our standard initial chemical abundances -  atoms except for 100\% H$_{2}$ - at the elemental abundances found in Table \ref{elem_ab}.  The fractional abundances obtained with these initial concentrations are also plotted in Fig.~\ref{CI_sens} with single black lines.   For most molecules, there is very little difference between the black lines and the distributions at any time, although what differences exist occur mainly at earlier times.  For complex molecules however, such as HC$_{7}$N, CH$_3$OH, and NH$_2$CHO, which typically peak at so-called early times in the range 10$^{4-6}$ yr, the early-time abundances are significantly higher with the standard than with the calculated initial abundances. 

One reason may lie in the relatively high abundance of atomic hydrogen calculated at steady-state in the diffuse stage.  Unlike the case for standard abundances, in which, starting from zero,  the fractional abundance of atomic H reaches a constant 10$^{-4}$ by 10$^{3}$ yr,  the calculated atomic hydrogen distribution at steady-state in the diffuse stage remains peaked at $\approx 3\times 10^{-3}$  for up to 10$^{5}$ yr in the dense cloud stage. Note that our model includes the H$_2$ and CO self-shielding using approximations from \citet{1996A&A...311..690L}.
 The excess atomic hydrogen calculated in the diffuse cloud stage and used initially in the dense cloud stage can interfere with the production of complex molecules by depleting some intermediate hydrocarbon radicals via neutral-neutral reactions. 
  Another possible reason is the relatively high abundance of CO calculated to occur at steady-state in the diffuse stage.  Once a significant amount of carbon becomes entrained in CO, it is not as easy to form complex carbon-containing species.  
 
To determine which of these effects is the more important for the dense cloud stage, we first varied the initial C$^{+}$/CO dense cloud ratio over a wide range
while fixing the overall carbon abundance as well as the other elemental abundances at the standard values shown in Table \ref{elem_ab}.  
   We separately did an analogous variation for the initial H/H$_2$ ratio in the dense cloud stage, in which case all the carbon lies initially in C$^+$.  Some results at a dense cloud time of 10$^{5}$ yr can be seen in Fig.~\ref{corr_init}.  Larger carbon-containing organic species are clearly enhanced for larger initial C$^+$/CO ratios, but there is a saturation point for C$^+$/CO larger than 1.  Because our standard initial  C$^{+}$/CO ratio exceeds unity  and our calculated dense cloud initial abundance ratio has a median value near 5,  the variation of this ratio between our two sets of initial concentrations  is {\em not} the reason for differences in abundances of complex molecules such as HC$_{7}$N. Regarding the initial fractional abundance of atomic hydrogen, however, we see that even  the smaller organic species CH$_{2}$ and HC$_{3}$N are reduced in abundance as the ratio increases to 10$^{-3}$; the reduction in abundance shown  is even larger for more complex species. Note that there is also a saturation effect here.   In particular, the molecular abundances are not sensitive to  H initial abundances smaller than $10^{-5}$, but this value is well below the range of values in our calculations. Within the ranges of initial concentrations for the dense cloud stage studied, the major source of sensitivity is thus the residual atomic hydrogen abundance.  For initial concentrations, however, in which there is more CO than C$^{+}$, the abundances of larger organic species will be significantly reduced at early times.  The fact that a large abundance of atomic hydrogen can significantly reduce calculated peak abundances of complex molecules at early time is of some importance considering that 21 cm absorption studies indicate significant abundances of atomic H in some  cold cores that are higher than calculated by our standard model \citep{2009AAS...21348509K}. This possible discrepancy argues against the high calculated early-time abundances for complex molecules in the simple pseudo-time-dependent model and for its possible replacement by models that take into account cloud formation and heterogeneity. On the other hand, the presence of anions or PAH's, not included in this study, tends to aid the synthesis of complex molecules \citep{2008ApJ...680..371W,2009ApJ...700..752W}.
 
 In the variations of the other parameters discussed below, we return to our standard initial chemical abundances for dense clouds.  This choice maximizes complex molecule abundances at early times.  Nevertheless, the abundances achieved are still low compared with low-metal elemental abundances, as can be seen in the recent calculations of \citet{2009ApJ...700..752W}. 

\subsection{Sensitivity to the reaction rate coefficients, temperature, and density}\label{Tn}

The sensitivities of atomic and molecular abundances in cold dense clouds to variations in reaction rate coefficients and gas temperature and density  have already been studied by \citet{2006A&A...451..551W}. For this reason, we will just mention the salient results of our new calculations.  For rate coefficients, the computed  abundances have log-normal distributions resulting from the probability distribution functions of the reaction rate coefficients \citep[see][]{2006A&A...451..551W}.   Moreover, a variation of temperature within the range $5-15$~K mainly affects the nitrogen-bearing species. In fact, much of the nitrogen chemistry starts with the reaction N$^+$ +H$_2$ $\rightarrow$ NH$^+$ + H, which is either slightly endothermic or possesses a small barrier.  The rate coefficient of this reaction at  temperatures under 100 K depends upon the fraction of molecular hydrogen in its excited ortho state; the results of more detailed calculations \citep{1991A&A...242..235L}  yield an effective activation energy of 85~K for the temperature range 10-100 K, as listed in the osu network. Temperatures below 10~K  then produce smaller abundances of  N-bearing species than temperatures above 10 K \citep[see Fig. 4 of][]{2006A&A...451..551W}. Within the chosen range, the density has little influence on the abundances.

\subsection{Sensitivity to the cosmic-ray ionization rate}

\begin{figure}
\begin{center}
\includegraphics[width=1\linewidth]{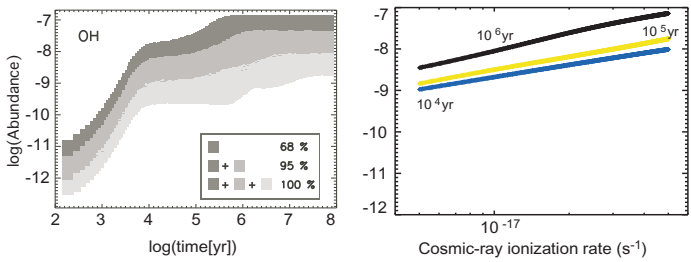}
\caption{OH sensitivity to the cosmic-ray ionization rate ($\zeta$). Left panel: probability density as a function of time for the different values of $\zeta$. Right panel: OH abundance as a function of $\zeta$ for three different times. \label{corr_zeta_OH}}
\end{center}
\end{figure}

\begin{figure}
\begin{center}
\includegraphics[width=1\linewidth]{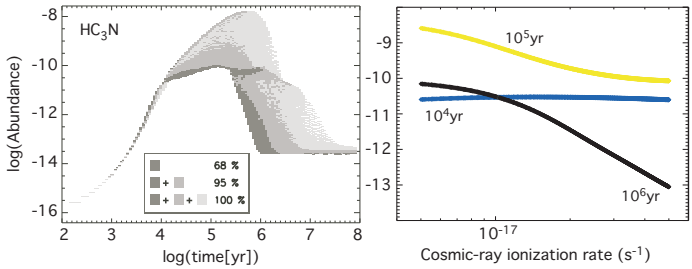}
\caption{HC$_3$N sensitivity to the cosmic-ray ionization rate ($\zeta$). Left panel: Probability density  as a function of time for the different values of $\zeta$. Right panel: HC$_3$N abundance as a function of $\zeta$ for three different times. \label{corr_zeta}}
\end{center}
\end{figure}

Most of the molecules are sensitive to the cosmic-ray ionization rate ($\zeta$) since cosmic ray ionization is the main source of atomic and molecular ions for cold clouds, and the abundances of neutral species are determined mainly by reactions involving ions and/or electrons. Sensitive species can be divided into two groups: 1) those  sensitive to a change in $\zeta$ at all times and 2) those only sensitive to $\zeta$ between $10^4$ and $10^7$~yr. The first group is composed of molecules that are more or less direct products of  the ionization of H$_2$ by cosmic rays, such as  H$_3^+$, OH (see Fig.~\ref{corr_zeta_OH}) and O$_2$. Complex molecules, such as the cyanopolyynes, are  part of the second group, as can be seen in Fig.~\ref{corr_zeta} for the case of HC$_{3}$N.  Some molecular abundances increase as $\zeta$ increases (for instance OH, CO, O$_2$, C$_3$, C$_3$O, SO, NO, and most of the ions), whereas others clearly decrease with increasing $\zeta$ (for instance the cyanopolyynes, H$_2$O, OCS, CS, CH$_3$OH and HCN). The cases of OH and HC$_{3}$N can be seen in the right panels of Figs.~\ref{corr_zeta_OH} and \ref{corr_zeta}. Presumably, the distinction for neutral species arises from whether or not ions are more important in the formation or the destruction of molecular species.  The species OH and H$_2$O for instance are both produced by the dissociative recombination of H$_3$O$^+$ but H$_2$O is mainly destroyed by an ion-molecular reaction with C$^+$ whereas OH is destroyed by reaction with neutral N.

 The dependence on $\zeta$ at steady state has been discussed previously by 
\citet{1996A&A...306L..21L} and \citet{1998A&A...334.1047L}, who found that when photodestruction by external photons is ignorable, then fractional abundances depend on the ratio $\zeta/n_{\rm H}$.  Thus the fact that the distribution for HC$_{3}$N collapses to  near a point at steady state despite the large variation in $\zeta$ shows that there is also very little dependence on $n_{\rm H}$ whereas the distribution for OH shows that this radical  has some (inverse) dependence on $n_{\rm H}$.

\subsection{Sensitivity to the elemental abundances}\label{EA}
 
 \begin{figure}
\begin{center}
\includegraphics[width=1\linewidth]{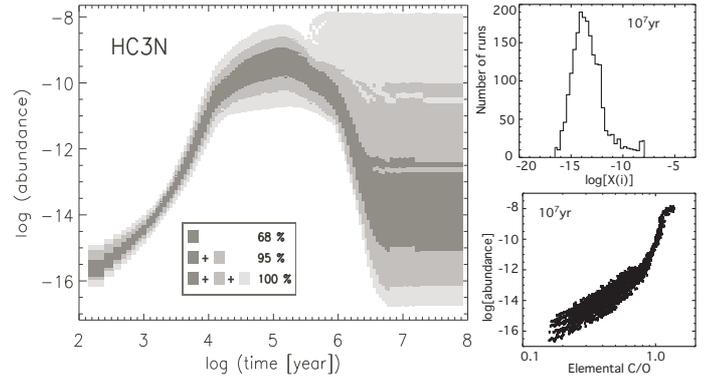}
\caption{HC$_3$N sensitivity to elemental abundances. Left panel: HC$_3$N probability density as a function of time. Right upper panel: distribution of HC$_3$N abundance at $10^7$~yr. Right lower panel: HC$_3$N abundance as a function of elemental C/O ratio at $10^7$~yr. \label{corr_AB}}
\end{center}
\end{figure}

The effects of variations in the  elemental abundances are shown in Fig.~\ref{corr_AB} for HC$_3$N, where the distribution spreads out with increasing time, reaching a standard deviation that exceeds 1.0, which implies a 68\% chance of finding the abundance to be spread over 2.5 orders of magnitude, when steady-state is finally reached.   The most important variations in the range considered are those for the elements carbon and oxygen, and most of the sensitivity is to the ratio of these elements, which varies between 0.2 and 1.4. The correlation between the HC$_3$N abundance at 10$^{7}$ yr and the C/O elemental ratio is shown in the right lower panel of the figure, where it can be seen that HC$_{3}$N increases in abundance quite dramatically with C/O  whereas the spread at a given C/O ratio caused by different values of C and O as well as variations of the other elemental abundances  is smaller.  In general, larger abundances of C-rich molecules such as HC$_3$N are obtained with increasing C/O. The reason is that the greater the C/O ratio, the greater the amount of carbon that is not taken up in the form of CO.  When the C/O elemental ratio becomes larger than 1, we go from a chemical regime dominated by oxygen to a regime where carbon chemistry is the more important.  Since it takes some time for a large amount of CO to be synthesized, the largest effect is seen at the longest times. 

\subsection{Comparison of the different sensitivities}

 \begin{figure}
\begin{center}
\includegraphics[width=1\linewidth]{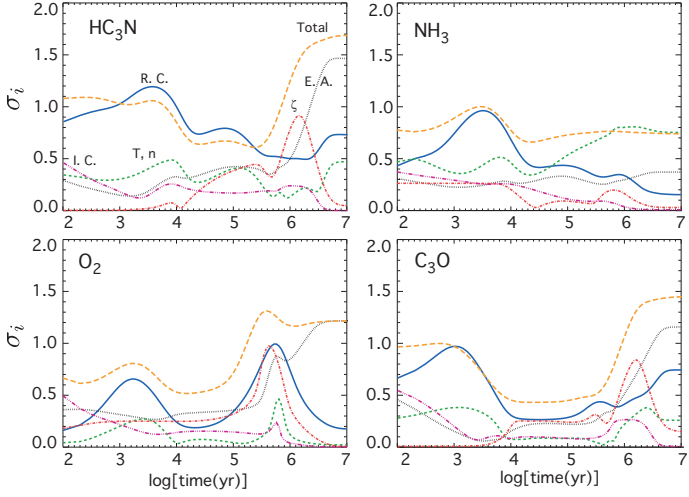}
\caption{Logarithmic standard deviations for different parameters $\sigma_{\rm i}$ as functions of time. R.C. stands for rate coefficients, I.C. for the initial concentrations, E.A. for elemental abundances, (T,n) for temperature and density, and $\zeta$ for cosmic-ray ionization rate. The long dashed line with designation All refers to the logarithmic standard deviation obtained by varying all parameters together. \label{spec_standdev}}
\end{center}
\end{figure}

 \begin{figure}
\begin{center}
\includegraphics[width=1\linewidth]{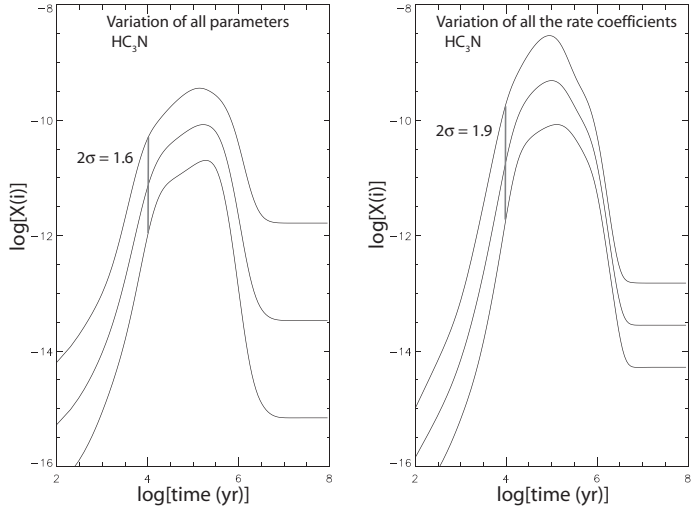}
\caption{Abundance of HC$_3$N as a function of time for the case of all parameters varied together on the left and for the variation of rate coefficient only on the right. In each panel, the middle curve represents the median abundance while the two external curves represent a 1$\sigma$ logarithmic standard deviation from this abundance. \label{standdev_HC3N}}
\end{center}
\end{figure}

 \begin{figure}
\begin{center}
\includegraphics[width=1\linewidth]{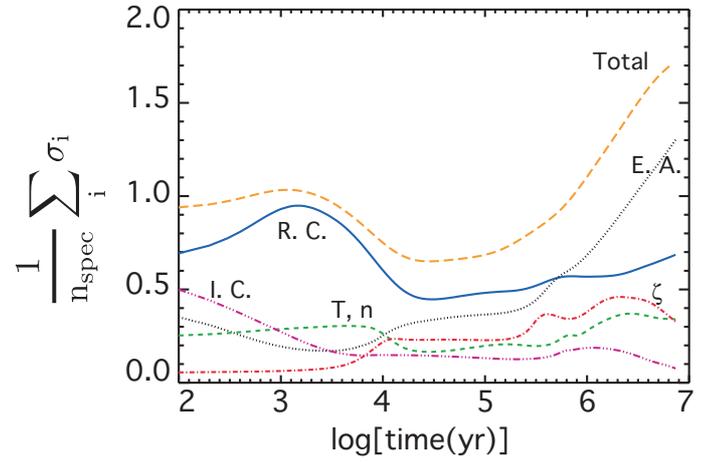}
\caption{Average logarithmic standard deviation over all species for each parameter or group of parameters.  The ordinate is the sum of the standard deviations of the individual molecular species divided by the number of molecular species.  \label{sum_standdev}}
\end{center}
\end{figure}

To compare the effect of each parameter, including the initial concentrations as described in Section~\ref{sens_ic},  we computed the logarithmic standard deviation $\sigma_{i}$  of the abundance distributions at each time for each parameter group $i$ -- initial concentrations, elemental abundances, cosmic-ray ionization rate, temperature/density, and reaction rate coefficients -- as well as for the case where we vary all parameters together.  Note again that  a logarithmic standard deviation of 1.0 refers to a factor of 10 in either direction (see Section~\ref{var_par}).  
As an illustration, we show the individual logarithmic standard deviations for several molecules as a function of time in Fig.~\ref{spec_standdev}.  Different molecules are sensitive to different parameters and this sensitivity varies with time. For HC$_3$N, the reaction rate coefficients  dominate until late times when the species is more sensitive to the elemental abundances (see Section~\ref{EA}). The species NH$_3$, on the other hand, is more sensitive to the gas temperature and density for times larger than $10^5$~yr (see Section~\ref{Tn}).  As expected, the standard deviation when varying all parameters   is generally larger than the ones for individual parameters but much smaller than the sum of the individual effects. 

In some cases, however, the standard deviation for all parameters is smaller than for one of the individual parameters or parameter groups, such as the rate-coefficient (``R. C.'')  case for HC$_3$N between $10^3$ and $10^5$ yr. This situation can happen because the amplification of the error affecting the rate coefficients  and its propagation into a spread of abundances depends on the set of parameters other than the rate coefficients. In some cases, the standard set of parameters will produce a dispersion that is larger than what is obtained with most of the other sets that are generated in the case for all parameters. As a consequence, the value of $\sigma(t)$ obtained by varying only the rate coefficients can be higher than what is found when varying all the parameters together, which attenuates the dispersion of abundances. Large values of $\sigma(t)$ are often associated with a stiff variation of the abundance, because one of the effects of changing the rate coefficients is to produce a time shifting of the chemical evolution. If a species experiences a stiff variation of its abundance at a time $t$, this time-shifting produced by varying the rate coefficients results in a large dispersion of its abundances at $t$. When the standard set of parameters produces the stiffest variation, the deviation in the R. C. case can exceed that for all of the parameters.  In Fig.~\ref{standdev_HC3N}, we show the mean abundance of HC$_3$N with its $\sigma$ standard deviation in the case of the variation of all parameters  and in the case of the variation of the rate coefficients  only. For each case, we show the standard deviation obtained at $10^4$ yr by a gray vertical line. In the  R.C case, the abundance of HC$_3$N increases much more stiffly than in the  case for all parameters and this introduces a larger standard deviation at a specific time. In this specific case, the change in stiffness, which also produces a difference of the mean abundance, arises because of the sensitivity of the stiffness to the initial conditions. As discussed in Section~\ref{sens_ic},  the  case of all parameters uses computed initial compositions whereas the R. C. case assumes the standard atomic composition, which results in a stiffer evolution around $10^4$ yr.  In the case of NH$_3$, as shown in Fig.~\ref{spec_standdev}, at $10^6$ yr  the sensitivity to the temperature  produces a slightly larger $\sigma$ than the case for all parameters at a time when the NH$_3$ abundance increases stiffly from $10^{-8}$ to $4\times 10^{-8}$ on a very short period of time. At that time and for the particular set of parameters, the stiffness of the  NH$_3$ abundance curve  is diluted by the variation of the other parameters.

To study the general effect on the complete model, we  calculated the average individual logarithmic standard deviations for all species as a function of time through 10$^{7}$ yr,  as can be seen in Fig.~\ref{sum_standdev}.    Although the relative importance of the parameters varies with time, sensitivity to the reaction rate coefficients dominates for ages less than $4\times 10^5$~yr. For later times, the elemental abundances are the main source of uncertainty in the model.  At early time, where model results are typically closest to observed values for cold cores, the average logarithmic standard deviations of the various parameters range from 0.5 (factor of  three; reaction rate coefficients) to 0.1 (factor of 1.3; initial concentrations).  Of course, these standard deviations are based on the ranges of the parameters.  Although most of these ranges are reasonable, based on a variety of criteria, the range for the elemental abundances does not cover some of the large differences between so-called high-metal and low-metal abundances.  With larger ranges, the sensitivity to elemental abundances is likely to increase in importance even at shorter times.  Another effect not included in the figure is that of the large difference between the range of initial H atom concentrations when the initial concentrations are varied, and the standard initial abundance of zero. Even the standard deviation introduced by a variation of initial H between 0 (standard initial abundance) and $10^{-3}$ is however smaller than all the standard deviations discussed in this section,  because its main effect is on the most complex molecules.

\section{Comparison with observations}

It is common to use chemical models to constrain some of the physical parameters of a cloud (cosmic-ray ionization rate, age, etc.) by comparing observed abundances with models in which the selected parameter is changed. Note that the age of a cloud is not one of the parameters that is handled by our random treatment; it is possible but not necessary to handle time randomly.  In any case, this method of constraint for one parameter or group of parameters requires that the other parameters be well known and/or that the model results, such as abundances, do not depend on the adopted values and uncertainties for the other parameters. Both hypotheses are usually not completely true, especially for reaction rate coefficients. Using our sensitivity analysis, we have previously studied the possibility of putting some constraints on some of the model parameters for dense clouds by comparing our modeling results with observations in two dense clouds L134N, with 42 species, and TMC-1, with 53 species \citep{2006A&A...451..551W}. Our major approach was to define agreement by the number of species for which the calculated distribution of abundances overlaps with the observational value and uncertainty.

To compare observed and modeled abundances here, we have instead computed at each time a function $D(t)$, defined by the equation  
\begin{equation}
 D(t) = \sum_j |{\rm log}(X^j_{\rm mod}(t))-{\rm log}(X^j_{\rm obs})|
\end{equation}
with $ X^j_{\rm obs}$ the observed fractional abundance of species $j$ and $\ X^j_{\rm mod}(t)$ the abundance of species $j$ computed by the model at time $t$.  For each parameter or group of parameters, we then attempted to determine the minimum value of $D(t)$ as a function of both the parameter and time. The smaller the value of $D$, the better the agreement. Our definition of D makes the ``agreement" with minor species such as cyanopolyynes as important as the agreement with major species such as CO. There is, in our opinion, no reason why CO should have more weight in the comparison than minor species. Indeed, reproducing the abundance of a minor species might indicate that the network works correctly at a more detailed level.  We could introduce a weighting factor if  we had some indication that some observed abundances are wrong or highly uncertain for instance. This is generally not the case here. Nevertheless, it is indeed true, as the referee hints, that the parameter $D$ is somewhat arbitrary.

\subsection{Constraints on the model parameters using all observed species}

\begin{figure}
\begin{center}
\includegraphics[width=1\linewidth]{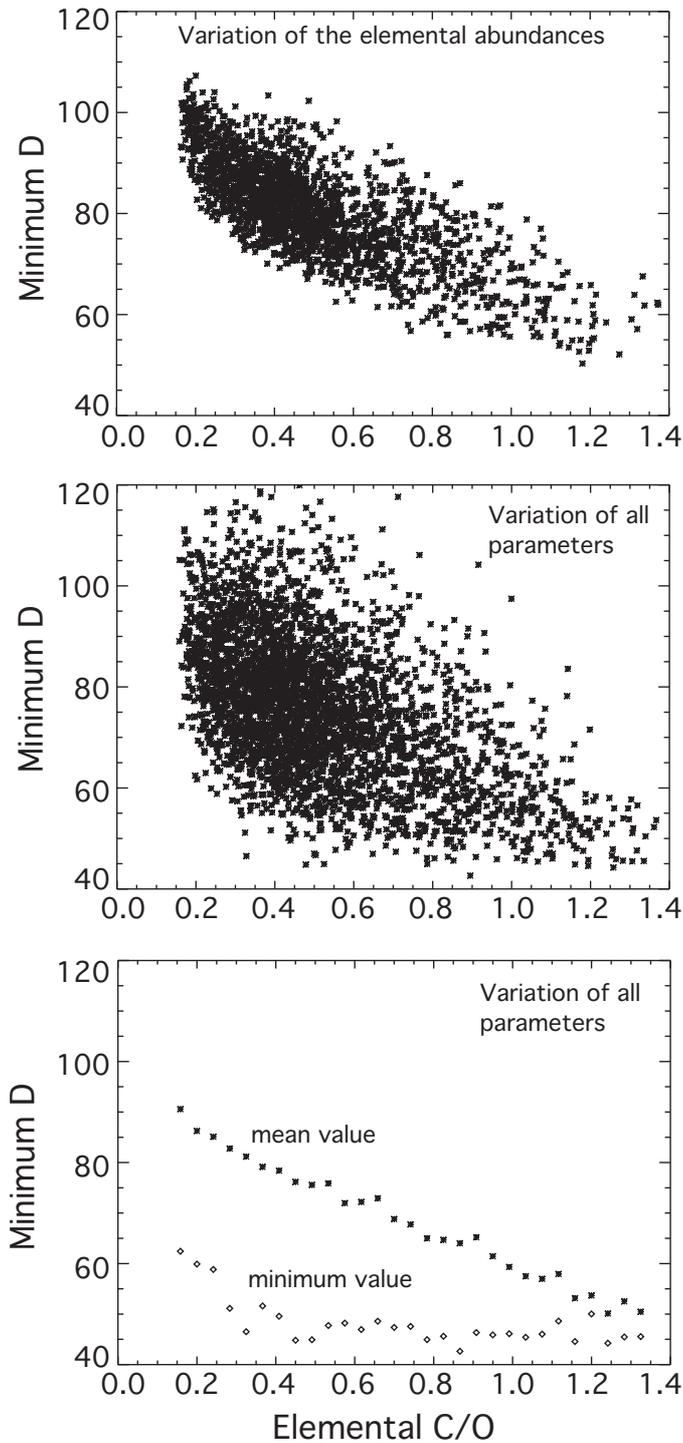}
\caption{Minimum D in TMC-1 as a function of the elemental ratio C/O. The left panel shows the results when we vary only the elemental abundances of the model, the middle panel shows the result when we vary all the parameters together, and the right panel shows the mean and minimum values from the middle panel. \label{agree_ABelem}}
\end{center}
\end{figure}

 Here we sum over all observed species to obtain $D$ for a given source.  An important result of our analysis is that the constraints on one specific parameter depend  on the values of the other parameters. This result is illustrated by Fig.~\ref{agree_ABelem}, which shows three separate plots.  On the left one, the minimum value of $D(t)$ for TMC-1 is displayed as a function of the elemental abundance ratio C/O for the case where we vary only the elemental abundances. The middle plot represents the same quantity but in the case where we vary all parameters together. As can be seen on this figure, there is a large spread for the minimum $D$ in the middle panel, and to observe the trend better, we show on the right plot, the mean values of minimum $D$ obtained in this case, as well as the minimum values as a function of elemental C/O. Although we see the same trend in both cases  that the observations in TMC-1 are better reproduced by larger values of C/O, we also see that  the best agreement ($D \approx 45$) also exists for smaller values of elemental C/O in the case where the other parameters are varied.
In other words,  the "best model" is not necessarily the one at the extreme value of a parameter. On the plots of the right of Fig.~\ref{agree_ABelem} for instance, the best agreement is rather flat from C/O=0.4 to 1.4. We can see this effect in other instances. For example,  if we only consider the agreement with the observations as a function of the initial ratio C$^+$/CO while keeping the elemental abundances  fixed and all the other elements  in the atomic form, we find that observations in TMC-1 are best reproduced if all the carbon lies in the atomic form. If we change other initial abundances, such as the initial H/H$_2$ and/or the initial N/N$_2$ abundance ratios, we find the best agreement for other values of the initial C$^+$/CO. 

We will not give here the set of parameters for the "best model" because we did not explore the full parameter space. Doing so would be too time-consuming. The best approach for the moment is to reduce the range of possible variation of the parameters, especially for the reaction rate coefficients, which seem to be dominant for many molecules.  

\subsection{Constraints on the model parameters using selected species}

\begin{figure}
\begin{center}
\includegraphics[width=1\linewidth]{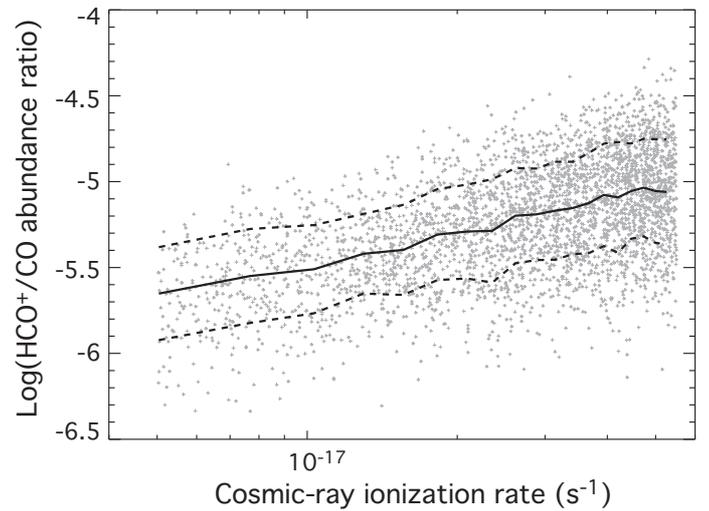}
\caption{HCO$^+$/CO abundance ratio as a function of $\zeta$ at $10^5$~yr in the case where we vary all parameters together. \label{HCOp_CO}}
\end{center}
\end{figure}

In the previous section, we  discussed some limitations on the use of the abundances of all  species observed in a cloud to constrain some of the model parameters via the method of minimization of $D$. One may do better by only using species sensitive to a given parameter so that their sensitivity is not masked by less sensitive species. Molecular ions, such as H$_3^+$, have been used to constrain the cosmic-ray ionization rate in diffuse regions \citep[see for instance][]{2003Natur.422..500M, 2004A&A...417..993L}.  In dense clouds, the abundance ratios HCO$^+$/CO and DCO$^+$/HCO$^+$ have been proposed to constrain the cosmic ray ionization rate and the fractional ionization \citep[see for instance][]{1998ApJ...499..234C}. We show in Fig.~\ref{HCOp_CO} the abundance ratio HCO$^+$/CO as a function of the cosmic-ray ionization $\zeta$ in the case where all parameters are varied at the same time. This ratio depends on time and Fig.~\ref{HCOp_CO} is for $10^5$~yr, the  so-called early time of best overall agreement by number of species. There is a clear correlation of the abundance ratio with $\zeta$ but the spread of possible HCO$^+$/CO values at a specific $\zeta$ is  at least one order of magnitude. According to \citet{2000ApJ...542..870D} and \citet{1992IAUS..150..171O}, the observed value of HCO$^+$/CO is about $1.5\times 10^{-4}$ in L134N and $10^{-4}$ in TMC-1 (CP). Compared with Fig.~\ref{HCOp_CO}, these results mean that  $\zeta$ should be $> 5 \times 10^{-17}$ s$^{-1}$ in  both clouds. The authors do not however give the uncertainties in the observed abundances.  Uncertainties should indeed be given to improve the utility of sensitivity studies.   The analysis of \citet{1998ApJ...499..234C} also indicates a rather high value of $\zeta$ in TMC-1(CP) in the range $(4-8) \times 10^{-17}$ s$^{-1}$.  If we plot the HCO$^+$/CO abundance ratio vs $\zeta /n$, a parameter discussed earlier, we obtain that ${\zeta /n}$ should exceed $ 5 \times  10^{-21}$  s$^{-1}$ cm$^3$.

\section{Discussion }

We have studied the sensitivity of the chemistry of cold dense cores to an assortment of parameters using the 0D gas-phase model Nahoon \citep{2005A&A...444..883W} in which chemical abundances evolve from given initial abundances. As variable parameters, we considered the temperature and density of the gas, the cosmic-ray ionization rate, the elemental abundances, the reaction rate coefficients, and initial concentrations.  These parameters {or groups of parameters} were all varied within realistic ranges of values either individually or simultaneously using Monte Carlo methods \citep{2004AstL...30..566V,2005A&A...444..883W,2006A&A...451..551W,2008ApJ...672..629V} used previously mainly to study the sensitivity of molecular abundances to chemical reaction rates.  For the case of the elemental abundances, our variations were carried out randomly from a base of abundances based on a new set first defined in \citet{2008ApJ...680..371W} as appropriate for cold dense clouds.  Our results depend to some extent upon whether individual parameters are varied while others are held fixed at standard values or whether all parameters are varied simultaneously.
Nevertheless, certain trends can be found.   Among the most important is the finding that, when averaged over all molecules,  the dominant source of uncertainty in predicted molecular abundances for times less than $4\times 10^5$~yr is the uncertainty in rate coefficients, whereas the dominant uncertainty at later times is caused by uncertainties in elemental abundances.  The former source of uncertainty can be reduced by improved laboratory and theoretical determinations of rate coefficients at appropriate temperatures, especially when sensitivity analyses point out the specific reactions of greatest importance, as done in \citet{2009A&A...495..513W}.  The latter source of uncertainty is astrophysical in origin, and suggests that more attention be paid to a physical understanding of how gas-phase elemental abundances evolve as dense clouds are formed from more diffuse gas.

Within the ranges shown in Table \ref{param}, the sensitivity to the other parameters is generally less. As long as the temperature and density of the gas uncertainties lie within their listed ranges, the molecular abundances predicted by the gas-phase model are quite robust; i.e., they do not depend much on $T$ and $n_{\rm H}$. The only exceptions are the N-bearing species containing only nitrogen and hydrogen, such as NH$_3$, which are produced much less efficiently at temperatures lower than the standard value of 10 K.  Many molecular abundances depend on the value of the cosmic-ray ionization rate, either at all times or only  between cloud ages of  $10^4$ and $10^7$~yr.  An example of the former is OH and one of the latter is  HC$_{3}$N, as shown in Fig.~\ref{corr_zeta}.  Of all the parameters, the initial concentrations with the chosen elemental abundances seem to be the parameter to which model results at all reasonable times are least sensitive.   Nevertheless, we found that cloud chemistry has to start with a significant fraction of the carbon in the atomic form and a very small amount of the hydrogen in atomic form to synthesize complex molecules.   Our standard model meets these requirements, but it may be in conflict with with some recent 21 cm studies by \cite{2009AAS...21348509K}, which show a larger amount of atomic hydrogen.  The range of atomic hydrogen abundances used when the concentrations are varied differs strongly from the standard model in which all hydrogen starts in its molecular form.  It is thus critical to understand both the physics and chemistry of the stages of collapse leading to the formation of cold cores for this problem as well as the elemental abundance problem. A new treatment of the gas-grain chemistry of cold cores in the process of formation is being prepared by Hassel, Herbst, \& Bergin.

In addition to the study of model sensitivity to assorted parameters, we have utilized the simultaneous variations of these parameters to attempt to determine their optimum values by comparison of calculated and observational results for the cold cloud cores TMC-1 and L134N.  In this paper, we used an indicator of agreement that minimizes the sum of the absolute values of the difference of the logarithms of the calculated and observed abundances. Constraints on individual model parameters are highly sensitive to the values of the other parameters (especially reaction rate coefficients) when variations are run simultaneously. It is currently not possible to constrain all the parameters by comparison with observed abundances by  rigorous methods so that efforts have to be made to reduce the uncertainties for some of the parameters. Uncertainties in rate coefficients for instance can be reduced by laboratory measurements or theoretical calculations on reactions clearly identified as quantitatively important for the model predictions.

Finally, although the strength of sensitivity methods for gas-phase chemical models has been clearly demonstrated in this and earlier studies for both cold clouds and more complex objects such as protoplanetary disks \citep{2008ApJ...672..629V},  the methods will need to be generalized to estimate the sensitivity to both chemical processes on grain surfaces and adsorption and desorption processes. The use of sensitivity methods for surface chemistry may become quite feasible within the next decade given the rapid pace of advance of both theory and experiments in this complex field.

\begin{acknowledgements}
We thank the referee for his careful reading of the paper and his suggestions.  E. H. acknowledges the support of the National Science Foundation (US) for the support of his program in astrochemistry,  the support of the NSF Center for the Chemistry of the Universe, and the support of NASA (NAI) for his work in the evolution of pre-planetary matter. V. W. and S. G. thank the French program PCMI for partial support of this work.
   
\end{acknowledgements}



\end{document}